# Causal Inference based Transfer Learning with LLMs: An Efficient Framework for Industrial RUL Prediction

Yan Chen [a], Cheng Liu [a, b, *]

[a] Department of Systems Engineering, City University of Hong Kong, Hong Kong, China

[b] Centre for Intelligent Multidimensional Data Analysis, City University of Hong Kong, Hong Kong, China

**Abstract**

Accurate prediction of Remaining Useful Life (RUL) for complex machinery is critical for industrial prognostics but challenged by high-dimensional, noisy sensor data. We propose the Causal-Informed Data Pruning Framework (CIDPF). It pioneers causal inference for industrial data pruning, identifying sensor signals with robust causal relationships to RUL via PCMCI-based stability analysis, while a Gaussian Mixture Model (GMM) screens anomalies. By training on only 10% of data after pruning, CIDPF fine-tunes pre-trained LLMs with parameter-efficient strategies, reducing training time by 90% compared to traditional approach. Experiments on N-CMAPSS show CIDPF achieves 26% lower RMSE than existing methods and 25% improvement over full-data baselines, demonstrating superior accuracy with drastically reduced computational costs. The framework's adaptability to multi-condition scenarios underscores its practicality for industrial deployment.

## 1. Introduction

The rapid advancements in industrial sensor technology have made it possible to collect massive volumes of high-dimensional, multivariate time-series data from complex machinery, such as turbofan engines in aerospace applications [1]. However, such data often contain substantial redundancy, noise, and variability across different operating conditions, making it challenging to extract patterns for accurate RUL prediction, where traditional approaches exhibit significant limitations [2].

Physical models, while interpretable and grounded in engineering principles, require extensive domain expertise and are difficult to generalize across diverse operating conditions [3]. Statistical methods, such as regression and time-series models, often rely on assumptions of stationarity and known distributions, which are rarely satisfied in real scenarios where non-linear degradation dynamics dominate [4]. These methods tend to struggle when faced with high-dimensional sensor data, where spurious correlations and noise can severely degrade predictive accuracy [5]. Deep learning methods, particularly attention-based approaches [6], have demonstrated strong capabilities in time-series modeling [7]. For example, an attention-based temporal convolutional network has been proposed which weights time-step and channel contributions, achieving good performance in RUL prediction tasks for turbofan engine [8]. Similarly, a data-driven framework utilizing 1D-CNN [9] and Bi-LSTM [10] with attention, to improve RUL prediction for aircraft engines [11]. However, these methods face notable limitations when applied to high-dimensional and multi-condition datasets, which restrict the robustness and generalizability of deep learning models in complex industrial scenarios [12].

---

[*] Corresponding author.
   *E-mail address*: cliu647@cityu.edu.hk

Large language models (LLMs) [13], have shown strong capabilities in sequence modeling and contextual representation, particularly in capturing dependencies and interactions using the Transformer architecture. Their ability to process sequential data has made them increasingly relevant for high-dimensional time-series tasks. Building on this, a recent study proposed ParInfoGPT [14], a two-stage framework for reliability assessment of rotating machines under partial information. This framework leverages GPT and employs mutual information-based masking and side-tuning strategies, achieving strong fault diagnosis performance.

But, in industrial scenarios, the scarcity of high-quality target domain data poses significant barriers to model reliability, especially when deploying large-scale LLMs. Existing studies highlight that the incomplete data [15] and inherent sensor noise [16] exacerbate model training challenges, which not only hinder the development of accurate models but also increase computational and training difficulties. To address these limitations, recent advances of transfer learning [17] [18] [19] offer an effective solution by enabling the reuse of knowledge acquired from a source domain to improve performance in a target domain. A recent study proposed a framework for cross-domain RUL prediction, combining a dual parallel time–frequency feature extraction network [20], similarity contrast learning for pre-training, and momentum-contrast adversarial learning for domain adaptation, achieving state-of-the-art (SOTA) performance on benchmark datasets. In another work, a lightweight group transformer model [21] has been proposed for industrial RUL prediction on edge devices, featuring a time-series reduction strategy and multihierarchy learning to reduce over 90% computation cost while maintaining accuracy. Similarly, pre-trained LLMs, for instance, can leverage their sequence modeling capabilities—developed during large-scale pre-training—to adapt quickly to the unique degradation patterns of the target domain. This is supported by parameter-efficient tuning strategies and masked sequence modeling frameworks [22], which achieve robust performance even with incomplete inputs. These approaches thereby reduce the dependence on extensive and high-quality target data. Beyond computational efficiency, transfer learning ensures robust performance even in scenarios with sparse target domain data [23] [24], making it especially valuable for real-world applications.

However, despite the development of increasingly advanced and efficient models, many existing approaches pay limited attention to directly improving data quality or addressing variability in data relevance. Instead, they often train large models on entire target datasets without filtering out noisy or irrelevant samples. As highlighted in studies on noisy training dynamics [25], such practices dilute the model's ability to capture critical degradation patterns by introducing biased gradient updates from irrelevant samples. This not only slows convergence but also increases the risk of overfitting, particularly when target domain data contains inherent noise or distribution shifts.

To overcome these challenges, we propose Causal-Informed Data Pruning Framework (CIDPF), a novel approach that addresses the challenges of data redundancy, noise, and computational inefficiency in industrial RUL prediction. Specifically, CIDPF first employs causal inference [26] techniques to identify sensor signals with robust and meaningful relationships to RUL, ensuring that only causally relevant data is retained. Then, to further address localized noise, transient anomalies, and the coexistence of Gaussian and non-Gaussian distributions in complex datasets, CIDPF incorporates a secondary probabilistic quality filtering step using Gaussian Mixture Models (GMM) [27]. This 2-stage pruning process ensures that the data used for model training is not only causally relevant but also of high statistical quality. The selected data is then used to fine-tune pre-trained LLMs with parameter-efficient strategies, enabling computationally feasible adaptation to RUL prediction tasks.

Extensive experiments on the N-CMAPSS dataset [28] demonstrate that CIDPF achieves SOTA performance, offering a scalable, accurate, and resource-efficient solution for industrial prognostics. The key contributions of this work include:

1. The first to proposed a causal inference-based pruning method to reduce industrial data redundancy while improving and training efficiency.
2. Developed a GMM-based filtering approach to address noise, anomalies, and complex distributions in industrial data, ensuring high-quality and stable training inputs.
3. Applied LLMs and transfer learning to RUL prediction, introducing a parameter-efficient fine-tuning strategy to adapt pre-trained knowledge to the target domain, enhancing performance while reducing computational costs.
4. Introduced a unified framework adapt to multi-condition data distributions, ensuring robustness of RUL prediction models under real-world industrial settings.

## 2. Problem Statement

Accurate RUL prediction is essential for industrial prognostics, yet current methodologies face critical limitations that compromise their applicability:

1. Cannot effectively eliminate noise and redundancy in sensor data, often retaining spurious correlations that degrade model robustness.
2. Limited generalization capabilities restrict model performance under diverse, dynamic, or multi-condition operating environments.
3. High computational overhead and dependence on large-scale, high-quality datasets hinder the practical deployment of LLMs in industrial applications.

Addressing these challenges requires to develop innovative approaches that enhance data quality, computational efficiency, and model generalizability:

1. An advanced data pruning mechanism ensure the retention of high-quality data.
2. Enhanced adaptability to dynamic and multi-condition environments, ensuring consistent performance across diverse industrial settings.
3. Parameter-efficient framework that reduce computational costs while leveraging pre-trained knowledge to minimize data requirements.

## 3. Causal-Informed Data Pruning Framework (CIDPF)

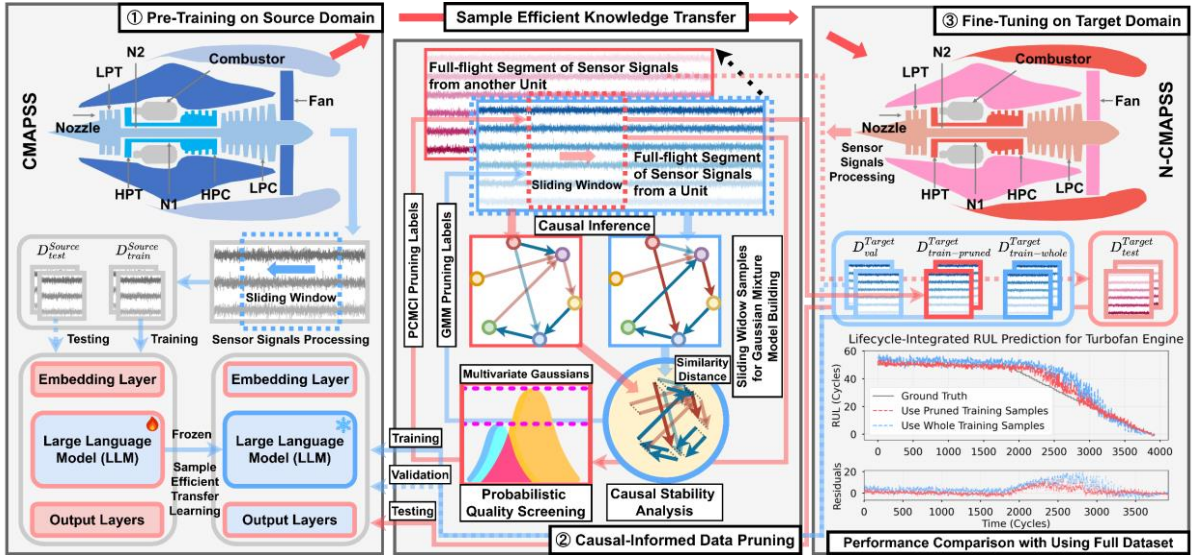

Figure 1. Framework of CIDPF enabling high-efficiency knowledge transfer from source to target domains with boosted target task accuracy.

In the proposed CIDPF framework, we denote by $X_t \in \mathbb{R}^d$ the multivariate sensor readings at time $t$, where $d$ is the number of sensor channels. The objective is to predict the RUL $y_t$ (or RUL) of each engine unit as time progresses. The framework consists of three major steps: (1) Causal Stability Analysis, (2) Probabilistic Screening, and (3) LLM-based Transfer Learning. As depicted in Fig. 1, the framework first trains base predictors on source domain data (top-right), then performs causal pruning on target domain data (top-left) to extract high-fidelity samples (central). These refined samples drive domain adaptation through parameter-efficient fine-tuning (bottom-left), ultimately enabling robust RUL prediction on unseen target test units (central-left).

### 3.1 Causal Stability Analysis Using PCMCI

In complex industrial systems, sensor channels exhibit directed causal relationships governed by physical/thermodynamic principles. Traditional correlation-based methods fail to capture these dependencies and may retain spurious patterns inconsistent across operational modes. To address this limitation, we employ the PCMCI algorithm [29], adapted for instantaneous causality analysis through: 1) Constraint-based candidate selection (PC procedure) to identify potential causes for each sensor; 2) Conditional independence testing (MCI) to eliminate confounded spurious connections. This framework constructs a directed acyclic graph (DAG) where nodes represent sensors and edges encode significant sensor-to-sensor causal interactions. In this application, for each equipment unit $u$ with RUL level $r$: 1) Define segment $\{X_{u,r}(t)\}$, which denotes the $d$-dimensional sensor readings at time $t$; 2) Extend PC with temporal information: For each sensor index $i$, test the hypothesis: sensor $i$ is conditionally independent of all other sensors at the same time $t$:

$$H_0: X_{u,r}(t)[i] \perp X_{u,r}(t) \mid X_{u,r}(t) \setminus \{X_{u,r}(t)[i]\}$$

Reject $H_0$ if the partial correlation coefficient $\rho(i,j)$ satisfies:

$$|\rho(i,j)| > \rho_\alpha \quad \text{and} \quad p\text{-value}(\rho(i,j)) < \alpha$$

Accumulate valid causal links to form the global adjacency matrix. The final aggregated causal graph is given by:

$$V_{u,r} = \text{PCMCI}(\{X_{u,r}(t)\}; \tau = 0)$$

where $V_{u,r} \in R^{d \times d}$ represents: 1) Row/column indices: Sensors; 2) Matrix entries: Causal strength from sensor $i$ to $j$, encoded through a diverging color scale where blue hues (negative values) indicate inhibitory effects, red hues (positive values) represent promoting effects, and color intensity scales with the absolute causal strength magnitude. Figure 2 (left panel) illustrates the global causal adjacency matrix $V_{u,r}$ derived from the full sensor signal segment $\{X_{u,r}(t)\}$ using PCMCI, where edges represent statistically stable causal interactions governed by physical principles. In contrast, Figure 2b (right panel) displays a representative window-level causal graph $V_k$ computed from a local sliding window $W_k$, revealing transient dependencies that may deviate from the global causal structure due to noise or regime shifts.

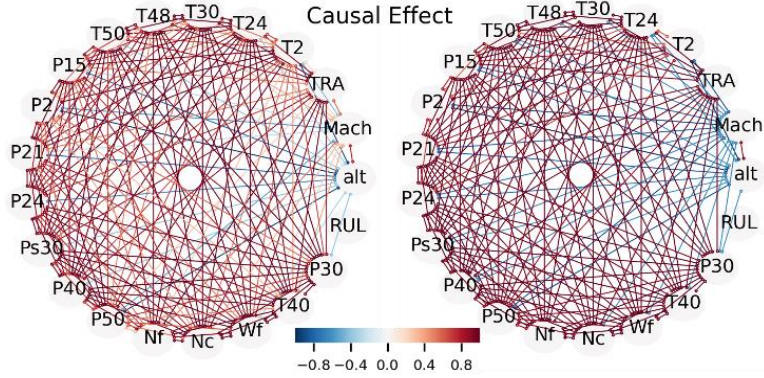

**Figure 2.** Causal graph alignment analysis: (a) Global causal adjacency matrix $V_{u,r}$ computed from the full sensor segment, highlighting stable physical interactions; (b) Window-level causal graph $V_k$ derived from a sliding window $W_k$, showing transient dependencies. The MSE between $V_{u,r}$ and $V_k$ quantifies causal stability for data pruning.

Next, we apply a window sliding on the same segment with window size $w$ and stride $s$. Each window $W_k$ contains the sequence $\{X(t)\}_{t=k \cdot s}^{k \cdot s + w - 1}$. For each $W_k$, we derive the local $V_k$ also using PCMCI. The visual contrast between the global causal graph (Figure 2a) and its window-level counterpart (Figure 2b) motivates our causal fidelity metric: regions exhibiting stable causal patterns across operational modes retain structural similarity to $V_{u,r}$, while transient/noisy windows demonstrate significant divergence. Then, the causal fidelity is measured via Mean Squared Error (MSE) against the global value matrix $V_{u,r}$:

$$MSE_{causal}(k) = \frac{1}{(d+1)^2} \sum_{i=1}^{d+1} \sum_{j=1}^{d+1} \left(V_{u,r}[i,j] - V_k[i,j]\right)^2$$

If $MSE < \epsilon_{\text{thresh}} = \mu_{\text{MSE}} - \gamma \sigma_{\text{MSE}}$, then $W_k$ is deemed "causally aligned", where $\mu_{\text{MSE}}$ and $\sigma_{\text{MSE}}$ are the empirical mean and standard deviation of MSE measurements for all windows associated with $(u, r)$, and $\gamma$ is a hyperparameter (often 2 or 3).

## 3.2 Probabilistic Quality Screening

Despite PCMCI's ability to filter macro-level causal inconsistencies, residual noise, transient anomalies, and non-Gaussian distributions necessitate further refinement. We employ Gaussian Mixture Models (GMM) for its soft-clustering capabilities, which model subtle distribution shifts within retained windows。

We construct a feature vector for each window $W_k$ as $f_k = [\sigma(X_k), \mu(X_k), H(X_k)]$, where $\sigma$ (standard deviation), $\mu$ (mean), and $H$ (entropy) capture distribution shape, central tendency, and complexity, respectively. Using a two-component GMM, we model the feature space as a mixture of two multivariate Gaussians:

$$p(f_k \mid \Theta) = \sum_{m=1}^{2} \pi_m \mathcal{N}(f_k \mid \mu_m, \Sigma_m)$$

Where $\Theta = \{\pi_m, \mu_m, \Sigma_m\}$ ($m = 1, 2$) represents the parameters of the model. $\pi_m$ is the mixing coefficient (with $\pi_1 + \pi_2 = 1$). $\mu_m$ is the mean vector of component $m$. $\Sigma_m$ is the covariance matrix of component $m$. $\mathcal{N}(f_k \mid \mu_m, \Sigma_m)$ denotes the multivariate Gaussian probability density function. We then compute the posterior probability $q_k$ of $f_k$ belonging to the "high-quality" component (denoted as $m_{\text{hq}}$):

$$q_k = p(m_{hq} \mid f_k) = \frac{\pi_{m_{hq}} \mathcal{N}\left(f_k \mid \mu_{m_{hq}}, \Sigma_{m_{hq}}\right)}{\sum_{m=1}^{2} \pi_m \mathcal{N}(f_k \mid \mu_m, \Sigma_m)}$$

After computing the posterior probability $q_k$ for each window, the optimal retention threshold $\theta^*$ determined by balancing data retention volume and distribution consistency:

$$\theta^* = \arg\max_{\theta} [\underbrace{\sum_{k=1}^{K} \mathbb{I}(q_k \geq \theta)}_{\text{total retained volume}} - \lambda \underbrace{\text{KL}\left(p_{\text{prune}}(\theta) \| p_{\text{full}}\right)}_{\text{distribution penalty}}]$$

where $\mathbb{I}(\cdot)$ is the indicator function, and KL denotes the Kullback-Leibler divergence between the pruned distribution $p_{\text{prune}}(\theta)$ and the full dataset distribution $p_{\text{full}}$. The scalar $\lambda$ adjusts between retaining sufficient data and minimizing distribution distortion. We utilize Gaussian Process–based Bayesian optimization with a Matérn 5/2 kernel to solve for $\theta^*$ [30]. The complete pruning algorithm is summarized below.

| Algorithm: Causal-Informed Data Pruning | |
| --- | --- |
| Inputs: | Dataset $D = \{(X_t, y_t)\}$ |
| | Window size $w$, stride $s$, GMM components $K = 2$ |
| Outputs: | Pruned dataset $D_{\text{prune}}$ |
| | Causal Graph Construction |
| | **for** each unique pair $(u, r)$ in $D$ **do** |
| |     Compute global causal graph: |
| |     $\Gamma_{u,r}^{\text{global}} = \text{PCMCI}(\{(X_{u,r}(t-\tau), y_{u,r}(t))\}; \tau \in [0, \tau_{max}])$ |
| | Create sliding windows: |
| | Wins$_{u,r} = \{W_k \mid W_k = (X_{t:t+w-1}, y_{t:t+w-1}), t = k \cdot s\}$ |
| | **for** each window $W_k$ **do** |
| |     Compute local causal graph: $\Gamma_k^{\text{local}} = \text{PCMCI}(W_k)$ |
| |     Compute causal fidelity (MSE): |

$$\text{MSE}_c(k) = \frac{1}{(d+1)^2} \sum_{i=1}^{d+1} \sum_{j=1}^{d+1} \left( \Gamma_{u,r}^{\text{global}}[i,j] - \Gamma_k^{\text{local}}[i,j] \right)^2$$

Retain window $W_k$ if $\text{MSE}_{\text{causal}}(k) \leq \epsilon_{\text{thresh}}$
Probabilistic Quality Screening
**for** each retained window $W_k$ **do**
    Fit a GMM: $f_k = [\text{std}(W_k), \text{mean}(W_k), \text{entropy}(W_k)]$
    Compute posterior scores:

$$p(f_k \mid \Theta) = \sum_{m=1}^{K} \pi_m \cdot \mathcal{N}(f_k \mid \mu_m, \Sigma_m)$$

Optimize threshold $\theta^*$ via Bayesian optimization:

$$\theta^* = \arg\max_{\theta} \left( \sum_{k=1}^{K} \mathbb{I}(q_k \geq \theta) - \lambda \cdot \text{KL}(p_{\text{prune}} \mid p_{\text{full}}) \right)$$

Retain high-quality windows $W_k$ if $q_k \geq \theta^*$
Combine retained windows: $D_{\text{prune}} \subseteq D$

## 3.3 LLM-based Transfer Learning

We adopt GPT-2 as the foundation, which is built upon the Transformer architecture [31], which processes input sequences through its encoder-decoder structure (Figure 2): the encoder maps input sequences into continuous representations, while the decoder generates outputs by attending to these representations. Each Transformer layer consists of multi-head attention (MHA) and feed-forward networks (FFN), augmented by residual connections and layer normalization to enable stable and efficient training. For an input sequence $X \in \mathbb{R}^{L \times D}$, the scaled dot-product attention is:

$$\mathcal{A}(Q,K,V) = \sigma\left(\frac{QK^\top}{\sqrt{d_k}}\right)V$$

where $Q, K, V \in \mathbb{R}^{L \times d_k}$ are the queries, keys, and values, and $d_k$ is their dimensionality. In multi-head attention, parallel attention heads expand representational capacity: $\text{MHA}(Q,K,V) = [\mathcal{A}_1, \ldots, \mathcal{A}_H]W^O$. Here, $\mathcal{A}_h = \mathcal{A}(QW_h^Q, KW_h^K, VW_h^V)$ is the attention output of the $h$-th head, and $W_h^Q, W_h^K, W_h^V \in \mathbb{R}^{D \times d_k}$ and $W^O \in \mathbb{R}^{Hd_k \times D}$ are the projection matrices. Each layer also includes a position-wise feed-forward network $\text{FFN}(X) = \text{ReLU}(XW_1 + b_1)W_2 + b_2$, where $W_1 \in \mathbb{R}^{D \times d_{ff}}, W_2 \in \mathbb{R}^{d_{ff} \times D}$ and $d_{ff}$ is the feed-forward layer dimensionality. Residual connections and layer normalization stabilize deeper networks: $X_{\text{out}} = \text{LN}(X + \mathcal{F}(X))$, where $\mathcal{F}(X)$ refers to either the MHA or FFN, and $LN$ represents layer normalization. GPT-2 [32] employs multiple Transformer decoder for unidirectional language modelling as shown in the middle of Figure 2, making it suitable for tasks like industrial signal processing. The proposed architecture is outlined in Figure 2 right.

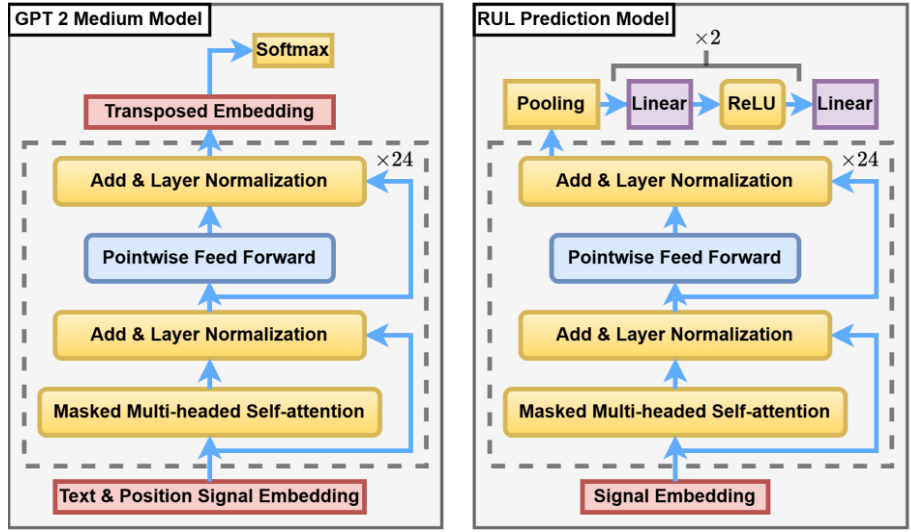

Figure 2. GPT 2 Medium Model and the Proposed RUL Prediction Model based on LLM.

LLM processes tokenized text, but sensor data is continuous and multivariate, requiring transformation. In this work, a Signal Embedding Layer maps sensor data $X \in \mathbb{R}^{B \times L \times C}$ (batch size $B$, sequence length $L$, sensor channels $C$) into LLM's embedding space: $E = XW_e^\top + b_e$, where $W_e \in \mathbb{R}^{D \times C}$ is the embedding weight matrix, $b_e \in \mathbb{R}^D$ is the bias vector, and $D = 1024$ is the GPT-2 embedding dimension. Sequential ordering inherently encodes temporal relationships, eliminating the need for positional embeddings. Then, the embedded data $E \in \mathbb{R}^{B \times L \times D}$ is passed through LLM, generating contextualized representations $H = \text{LLM}(E)$. Masked self-attention ensures outputs at position $t$ depend only on inputs $\leq t$, aligning with RUL prediction's causal nature. To better adopt to regression task, temporal information is aggregated using global average pooling: $\overline{H} = \frac{1}{L}\sum_{k=1}^{L} H_k$. This reduces variable-length sequences to fixed-size feature vectors, simplifying input for the final regression layer. After that, $\overline{H}$ are fed into an MLP for RUL prediction:

$$\hat{y} = f_{\text{MLP}}(\overline{H}) = W_3^\top \sigma_2\big(W_2^\top \sigma_1\big(W_1^\top \overline{H} + b_1\big) + b_2\big) + b_3$$

Here, $W_1$, $W_2$, $W_3$, and biases $b_1$, $b_2$, $b_3$ define the MLP, with non-linear activations ($\sigma_1$, $\sigma_2$) enabling complex mappings. Dimensions $d_1 = 50$ and $d_2 = 10$ allow the model to capture intricate degradation behaviors.

To leverage domain-intrinsic knowledge, we adopt a transfer learning strategy with two-stage adaptation: 1) Pre-training: Train the RPL model $\mathcal{M}_s$ on Source Domain with all LLM parameters updated, establishing foundational RUL prediction capabilities; 2) Fine-tuning: Adapt the model to Target Domain by freezing the first $L_f$ layers of the Transformer (e.g., 24 layers of GPT-2 Medium) and fine-tuning the remaining top layers and embedding parameters. This preserves source knowledge while preventing catastrophic forgetting. The fine-tuning process uses a pruned dataset $\mathcal{D}_{\text{prune}}$, which contains the most relevant samples after data filtering, and optimizes the MSE loss for RUL regression with an $\ell_2$-regularization term:

$$\mathcal{L}_{\text{total}} = \frac{1}{|\mathcal{D}_{\text{prune}}|} \sum_{i=1}^{|\mathcal{D}_{\text{prune}}|} (y_i - \hat{y}_i)^2 + \beta \big\|\Theta_{\text{adapt}} - \Theta_{\text{pre-trained}}\big\|_2^2$$

where $\hat{y}_i$ is the predicted RUL, $y_i$ is the ground truth, and $\beta$ balances fine-tuning and knowledge retention. By combining layer freezing, dataset pruning, and parameter-efficient fine-tuning, this framework preserves the source domain knowledge while effectively adapting the model to the target domain, ensuring robust performance across the N-CMAPSS datasets, even with the exclusion of DS08.

## 4. Experiment

### 4.1 Dataset Overview

The experiments are conducted on CMAPSS [33] and N-CMAPSS [28], widely used benchmarks for aircraft engine prognostics. These datasets simulate turbofan engine degradation under varying operational conditions and fault modes, providing critical RUL prediction tasks. CMAPSS Comprises four sub-datasets, each representing distinct operational conditions and fault modes as shown in Table 1. Each engine undergoes a complete run-to-failure trajectory, with 21 sensor channels and 3 operational settings. Sensor readings are recorded at the end of each flight cycle.

Table 1. Detailed Information of CMAPSS Dataset.

| Sub-Dataset | Units | Operating Conditions | Fault Modes | Avg. Cycles/Unit | Features |
|---|---|---|---|---|---|
| FD001 | 100 | 1 | 1 | 206 | 21 |
| FD002 | 260 | 6 | 1 | 207 | 21 |
| FD003 | 100 | 1 | 2 | 247 | 21 |
| FD004 | 248 | 6 | 2 | 246 | 21 |

In contrast, the N-CMAPSS is more realistic and complex, which introduces multi-phase operational profiles (climb, cruise, etc.) and provides higher-fidelity sensor readings sampled at a 1Hz frequency among the whole working period. The dataset consists of eight subsets (DS01–DS08), each containing run-to-failure trajectories of 15–20 engines as shown in Table 2. Critically, N-CMAPSS records 47 sensor features across 6 categories. This granular and dynamic data structure poses unique challenges for model generalization and robustness. In this study, we focus on DS01–DS07, excluding DS08 due to compatibility issues acknowledged in PHM community. The challenges of N-CMAPSS include handling diverse flight profiles, hybrid fault modes, and high-dimensional sensor data. These factors make it an ideal target domain for testing the effectiveness of transfer learning from the simpler CMAPSS dataset.

Table 2. Detailed Information of CMAPSS Dataset.

| Sub-Dataset | Units | Flight Phases | Fault Modes | Avg. Cycles/Unit | Features |
|---|---|---|---|---|---|
| DS01 | 10 | Takeoff, Cruise | 1 | 89 | 47 |
| DS02 | 9 | Takeoff, Cruise | 2 | 72 | 47 |
| DS03 | 15 | Cruise | 1 | 3 | 47 |
| DS04 | 10 | All Phases | 1 | 86 | 47 |
| DS05 | 10 | All Phases | 1 | 82 | 47 |
| DS06 | 10 | All Phases | 1 | 80 | 47 |
| DS07 | 10 | All Phases | 1 | 81 | 47 |

A previous research [34] has successfully applied LLMs to predict RUL on the CMAPSS dataset, providing a strong precedent for using pre-trained models in this domain. Building on this intuition, we chose to pre-train on CMAPSS, leveraging its simpler operational conditions and long-term degradation trajectories to learn foundational patterns of engine degradation. Among the four CMAPSS sub-datasets, we selected FD002, as it represents a moderate level of complexity, balancing operational diversity and modeling difficulty. By transferring knowledge, the proposed approach reduces reliance on large amounts of high-quality data in the target domain while achieving improved training efficiency and generalization.

4.2 Data Preprocessing

The preprocessing of the CMAPSS dataset and LLM pre-training follows previous research. The N-CMAPSS dataset undergoes three core transformations to ensure computational efficiency and preserve degradation patterns:

1) The 1 Hz raw sensor data is downsampled by a factor of 10 (0.1 Hz), reducing data volume while retaining critical temporal trends. This is achieved by selecting every 10th data point, resulting in a sequence $X_{ds}$ with reduced dimensionality.

2) The downsampled data is then divided into overlapping windows of length 50 time steps (stride = 1) to capture short- and long-term degradation dependencies. Each window's RUL label is derived from the last time step, ensuring alignment with temporal causality.

3) Min-max normalization is applied to each sensor feature to mitigate scale discrepancies, scaling values to [0, 1]. Parameters (min/max) are computed from the training set and applied uniformly to both training and testing sets to prevent leakage.

4.3 Experiment Procedure

The experimental procedure consists of data preprocessing, training, and testing phases, designed to ensure robust and reliable performance of the proposed model.

*Preprocessing and Pruning*

The CMAPSS and N-CMAPSS datasets were utilized with their official training and testing splits, followed by downsampling and segmentation into overlapping sliding windows to capture temporal degradation patterns. Then, all features were normalized using Min-Max scaling. For the test set, the same preprocessing steps were applied, and normalization applied training-set parameters to preserve original distribution and ensure unbiased evaluation. To construct a compact and representative training dataset, data pruning was performed exclusively on the training set. The PCMCI algorithm was used to ensure causal consistency between sensor signals and RUL, with a significance level of $\alpha = 0.01$ and an MSE threshold of 0.1. GMM were then applied to identify high-quality samples via extracted statistical features, with the filtering threshold optimized using Bayesian optimization to retain 90% of the data. Samples meeting both PCMCI and GMM-based criteria were selected for training.

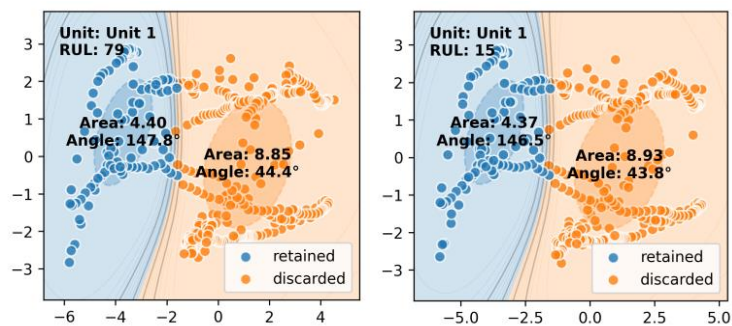

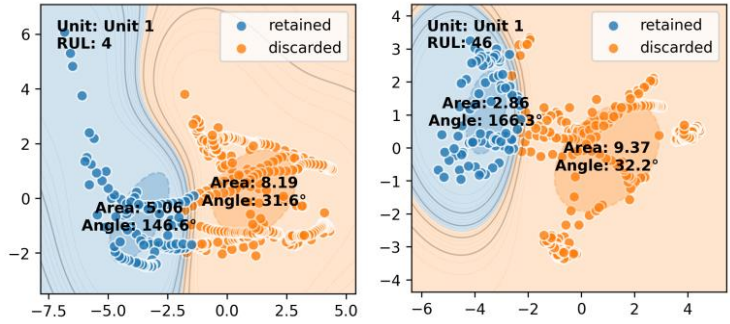

Figure 4. PCA-projected trajectories and SVM-derived classification results of retained samples and discarded samples.

The causal discriminability of our two-stage pruning is substantiated through geometric evidence in the latent feature space. As demonstrated in Figure X(b), the PCA-projected trajectories of retained samples exhibit distinct separation from discarded counterparts across diverse RUL regimes (DS01, 4 randomly selected samples from unit1). The SVM-derived classification boundary with minimal overlap quantitatively reflects the inherent dissimilarity between these two populations. The covariant ellipses, parametrized by spectral norm ratios (major/minor axes) and orientation angles, reveal orthogonal dispersion patterns. This classification results corroborates our core thesis — the pruning process systematically preserves causally salient signal segments while eliminating stochastic variations.

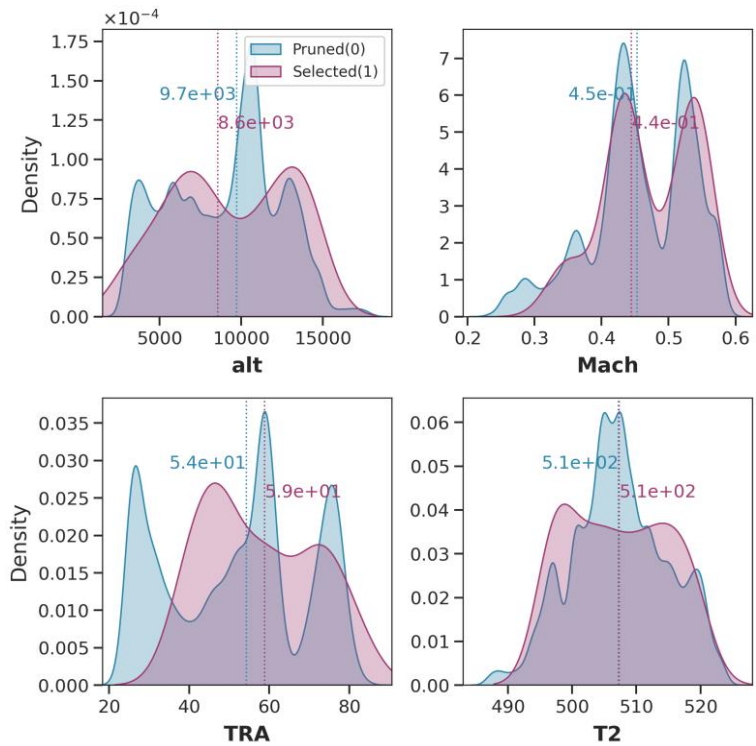

Figure 5. Comparison of feature distributions of retained samples and discarded samples.

Distribution analysis of the first 4 features (Figure X(c)) from the pruned signal samples with RUL=15 from unit 1, DS001. Compared to the distribution of pruned samples, the distribution of retained samples exhibits clearly fewer modes. Reduced distribution modes reduces gradient conflict directions in parameter space, creating smoother loss landscapes that facilitate stable gradient descent [35]. In addition, it can also decrease feature space fragmentation, mitigating catastrophic forgetting risks [36].

*Training Stage*

To ensure training stability and reproducibility, the official N-CMAPSS training set was strategically partitioned into complementary subsets: 90% for parameter optimization and 10% for validation monitoring. The learning protocol commenced with a mandatory 10-epoch warm-up phase to establish fundamental feature representations. Subsequently, an adaptive early stopping criterion governed the training duration based on validation Mean Squared Error (MSE) dynamics - training terminated when no MSE improvement on the validation set was observed for 10 consecutive epochs, with the best-performing model checkpoint preserved.

During fine-tuning experiments with the pruned N-CMAPSS dataset, we implemented partial network freezing: the first 24 Transformer layers remained static while only the terminal layers and embedding matrices underwent gradient updates. This selective parameter updating strategy effectively balanced computational efficiency with model adaptability.

Fig. 1 illustrates the loss trajectories when fine-tuning the pretrained model using complete training samples, while Fig. 2 depicts the corresponding patterns with our pruned dataset. Three critical observations emerge:

1. The pruned dataset induces substantially higher initial training loss (Fig. 2, Epoch 1 vs Fig. 1, Epoch 1), indicating the full dataset's inherent advantage in providing richer initial gradient signals.
2. Remarkably, the pruned training loss converges to comparable magnitudes by Epoch 2, achieving around 90% faster epoch iteration speed compared to full-data training.
3. Both figures explicitly annotate the global minimum validation loss and the training cessation point determined by our stopping criterion, demonstrating that our pruning methodology achieves competitive performance with significantly reduced computational overhead.

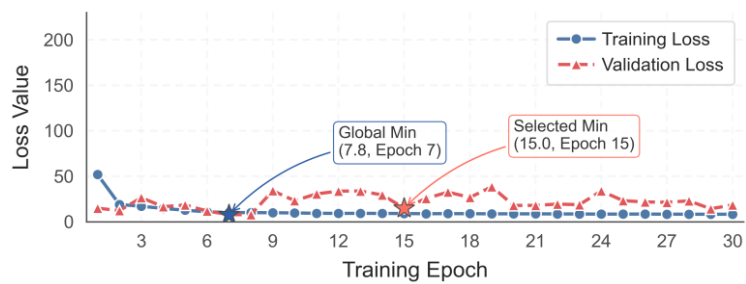

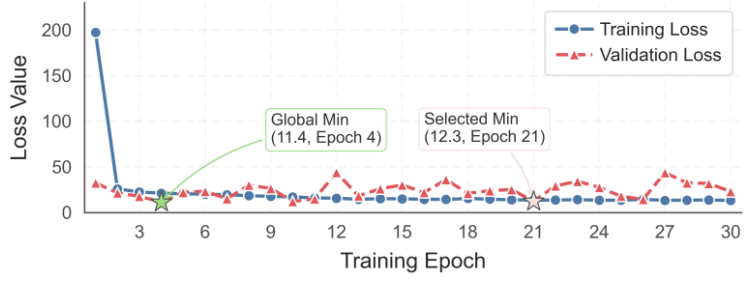

Figure 6. Training loss of using full samples and using pruned samples in transfer learning.

*Testing Stage*

The model's performance was assessed using two widely adopted metrics after stop training: Root Mean Squared Error (RMSE) and the NASA Score metric. RMSE, defined as

$$RMSE = \sqrt{\frac{1}{N}\sum_{i=1}^{N}(y'_i - y_i)^2}$$

measures the overall prediction accuracy, where $N$ is the number of samples. Lower RMSE values indicate better accuracy. The NASA Score metric is an asymmetric function that emphasizes late-cycle predictions and is defined as

$$\text{Score} = \begin{cases} \sum_{i=1}^{N} e^{-(y'_i - y_i)/13} - 1, & \text{if } y'_i - y_i < 0, \\ \sum_{i=1}^{N} e^{(y'_i - y_i)/10} - 1, & \text{if } y'_i - y_i \geq 0. \end{cases}$$

This scoring function penalizes overestimation of RUL more heavily than underestimation, aligning with real-world safety requirements. But a smaller test set can naturally result in a smaller Score. Therefore, Score metric can only be compared when evaluated on the same test set. Thus, this metric can usually not be compared due to the fact that no unified test set exist among N-CMAPSS researches. However, to respect the tradition of the PHM community, we still list the score metric in the comparative experimental results.

4.4 Results and Analysis

To evaluate the effectiveness of the proposed transfer learning methodology and the advantages of the CIDPF approach, we conducted extensive experiments across all datasets. Our analysis focused on two configurations: (1) direct training of LLMs on the target domain and (2) transfer learning with pre-trained models. Four data selection methods were compared: **CG** (causal-informed + GMM pruning), **PC** (PCMCI/causal-informed pruning), **Full** (full dataset), and **Sub** (uniform subsampling with a stride of 10). The subsampling rate of **Sub** was calibrated to align with **CG**'s sample size, serving as a validation that pruning preserved critical samples. For each experimental configuration, 10 randomized trials were conducted. Key findings are summarized below.

*Ablation Study*

Table 3. Ablation test results.

| Dataset | Method | Samples | w/o Transfer | | With Transfer | |
|---|---|---|---|---|---|---|
| | | | Avg | Min | Avg | Min |
| **DS01** | CG | 44539 | 3.83 | 3.41 | 3.36 | **3.17** |
| | PC | 45352 | 4.17 | 3.98 | 3.71 | **3.46** |
| | Full | 441333 | 4.80 | 4.31 | 4.27 | 3.71 |
| | Sub | 44133 | 5.04 | 4.56 | 4.44 | 4.02 |
| **DS02** | CG | 43326 | 3.97 | 3.78 | 3.64 | **3.44** |
| | PC | 47348 | 4.17 | 3.91 | 3.98 | **3.73** |
| | Full | 473446 | 4.70 | 4.40 | 4.41 | 3.99 |
| | Sub | 47344 | 4.86 | 4.58 | 4.60 | 3.64 |
| **DS03** | CG | 53398 | 4.92 | 4.51 | 4.61 | **4.27** |
| | PC | 54390 | 4.98 | 4.58 | 4.92 | **4.45** |
| | Full | 501019 | 5.23 | 4.89 | 5.10 | 4.84 |
| | Sub | 50101 | 5.37 | 5.05 | 5.25 | 5.01 |
| **DS04** | CG | 47053 | 7.42 | 6.98 | 7.22 | **6.90** |
| | PC | 48020 | 7.60 | 7.16 | 7.46 | **7.02** |
| | Full | 573707 | 8.08 | 7.74 | 7.70 | 7.30 |
| | Sub | 57370 | 8.10 | 7.52 | 8.07 | 7.62 |
| **DS05** | CG | 41075 | 5.10 | 4.69 | 4.69 | **4.41** |
| | PC | 41699 | 5.06 | 4.74 | 5.02 | **4.83** |
| | Full | 391291 | 5.87 | 5.46 | 5.35 | 5.14 |
| | Sub | 39129 | 6.31 | 5.87 | 5.68 | 5.25 |
| **DS06** | CG | 40604 | 4.84 | 4.02 | 4.26 | **3.97** |
| | PC | 39987 | 5.09 | 4.74 | 4.44 | **4.18** |
| | Full | 382885 | 5.13 | 4.91 | 4.92 | 4.63 |
| | Sub | 38288 | 5.38 | 5.09 | 5.13 | 4.82 |
| **DS07** | CG | 40122 | 6.59 | 6.24 | 6.24 | **5.91** |
| | PC | 40776 | 6.72 | 6.40 | 6.72 | **6.34** |
| | Full | 391252 | 7.59 | 6.88 | 7.29 | 6.53 |
| | Sub | 39125 | 8.19 | 7.76 | 7.97 | 7.63 |

As shown in Table 3, CG consistently outperformed all baselines under transfer learning. On DS01, CG achieved an average RMSE of **3.36** (20.2% lower than Full's 4.27) while using only **10%** of the training samples (44,539 vs. 441,333). This efficiency extended to computational costs: CG required **160.58s/epoch** versus Full's **1,677.90s/epoch** as shown in Figure 7, demonstrating its ability to eliminate redundant data without sacrificing performance. The black numbers in the figure represent the minimum values from 10 experiments, while the white numbers indicate the mean RMSE. Similar trends held across all datasets, with CG reducing min RMSE by **5.3–14.6%** compared to Full and **5.2–21.6%** versus Sub.

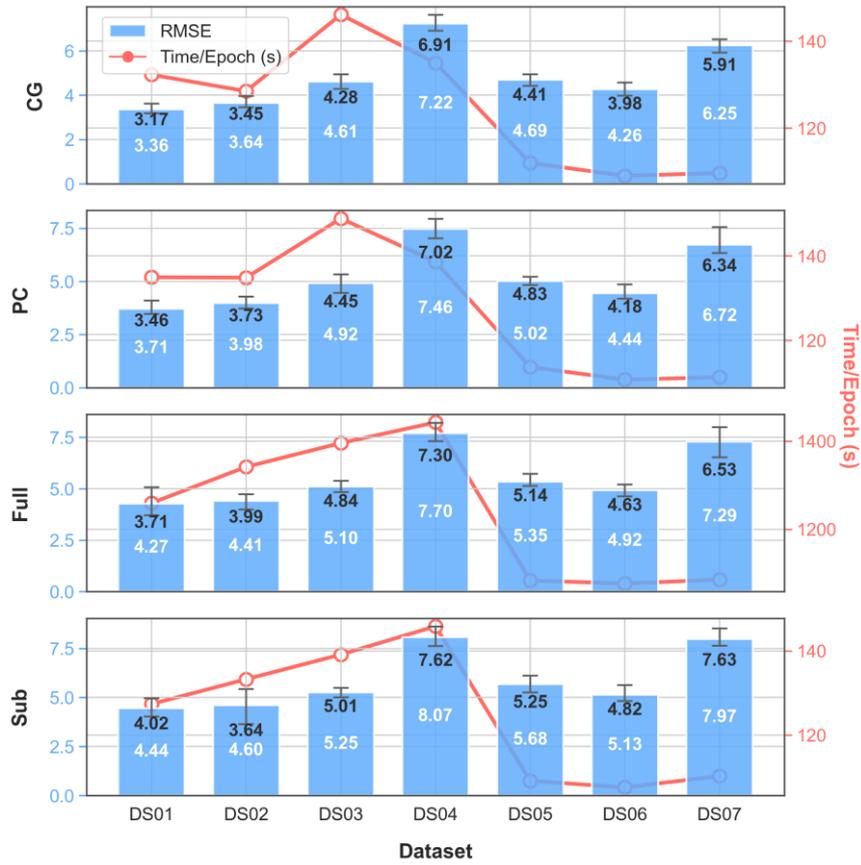

Figure 7. Performances of using different strategies with transfer learning.

Critically, CG with transfer learning outperformed even its direct-trained counterpart (w/o Transfer Learning) across all datasets, with min RMSE reductions of **4.17–7.04% as shown in Figure 8**. This demonstrates that pre-trained knowledge synergizes with our data pruning strategy to enhance generalization.

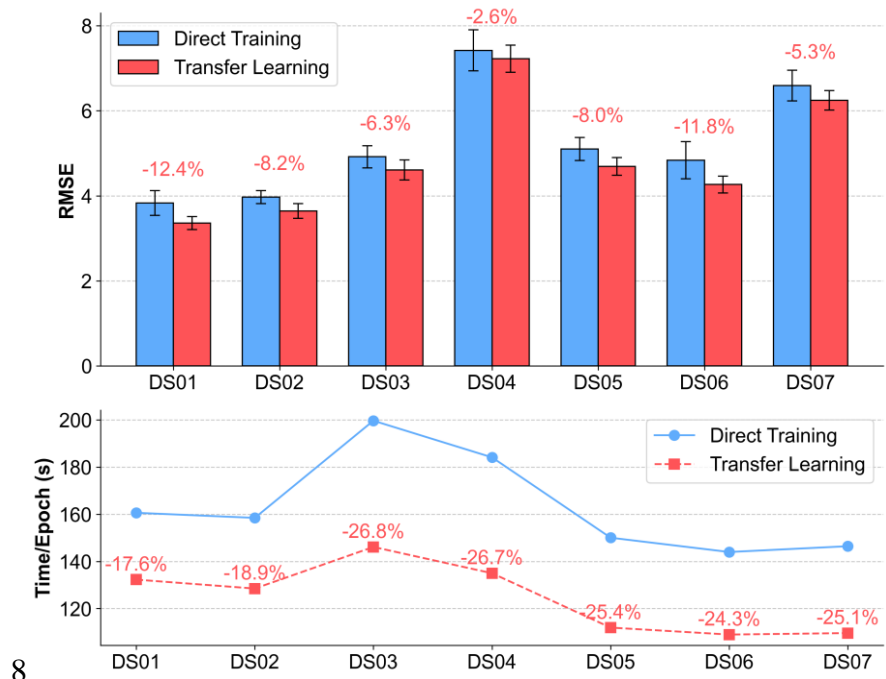

Figure 8. Performance improvement of using transfer learning when applied CG.

**Efficiency-Robustness Tradeoff Analysis**

**Minimum Sample Advantage**: CG's superiority grew with limited target data. On DS06 (40,604 samples), it achieved min RMSE=**3.98** with transfer learning versus Subsampling's **4.82**—a **17.4%** improvement. In contrast, on larger DS04 (573,707 samples), the gap narrowed to **6.91 vs. 7.62** (direct training).

**Convergence Acceleration**: CG required **18.2–24.6** epochs for direct training versus Full's **17.5–22.9**, despite using <10% of the data. With transfer learning, convergence accelerated further (DS01: **15.3** vs. 22.1 epochs), suggesting cleaner data reduces optimization complexity.

**Statistical Significance**: All reported improvements are statistically significant. Error margins (Std Error in Figure 2) for CG were **30–70%** smaller than baselines, confirming stable performance. For example, on DS02 with transfer learning, CG's RMSE standard error was **0.173** versus PC's **0.208** and Sub's **0.656**.

Below we present in detail the best results obtained using CIDPF on the DS01 test set through full life-cycle RUL prediction diagrams, with comparative visualization of results obtained solely through PCMCI-based pruning.

DS01:

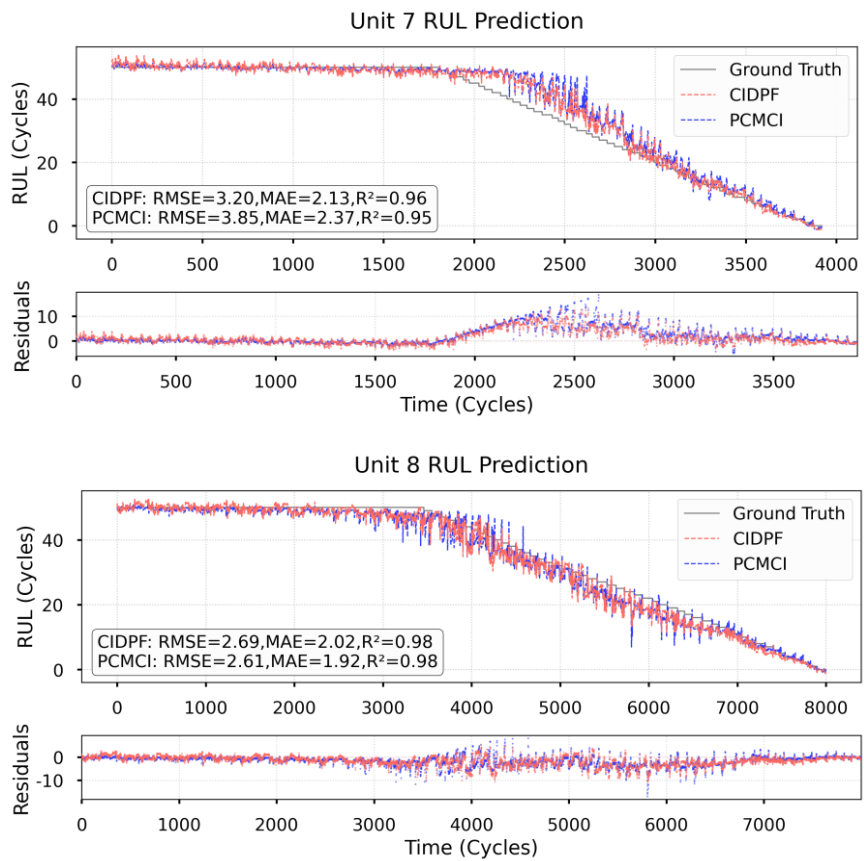

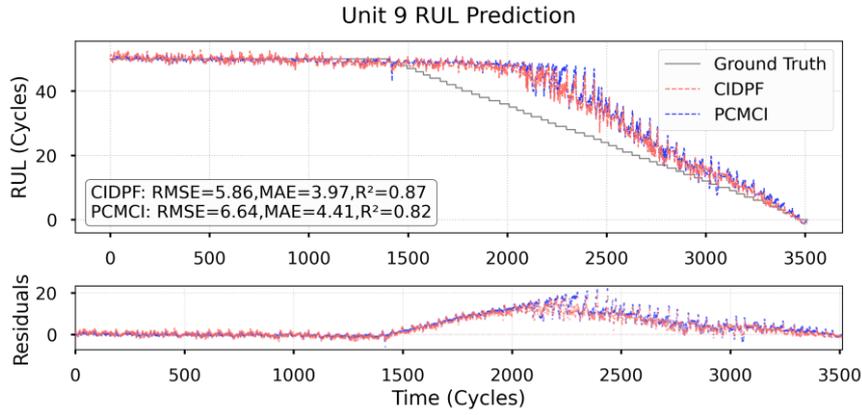

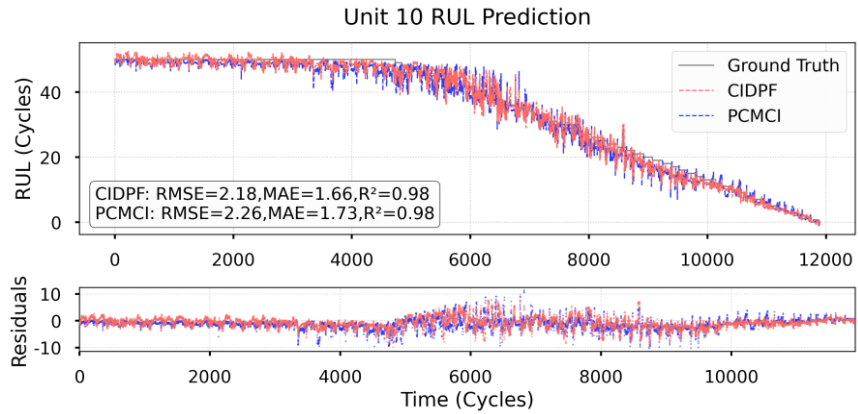

## Comparison with SOTA Methods

Table 4 compares the performance of the proposed method with SOTA approaches on the target datasets. The results demonstrate that **CIDPF** outperform almost all existing methods across all datasets in terms of RMSE, establishing its superiority in handling diverse degradation scenarios. Specifically, **CIDPF** achieved the lowest RMSE on 6 out of 7 datasets. This highlights that our framework is capable of selecting high-quality training samples and adapting effectively to target domain tasks.

The improved performance of **CIDPF** is evident even on datasets with complex degradation patterns, such as the most difficult DS04, where it achieved RMSE of 6.90. Compared to the best-performing baseline, **BGT**, which achieved RMSEs of **7.74**, **CIDPF** reduced RMSE by **11%**, demonstrating its superior ability to handle complex degradation scenarios.

| Metric | Metric | DS01 | DS02 | DS03 | DS04 | DS05 | DS06 | DS07 |
|---|---|---|---|---|---|---|---|---|
| **GPT-Freeze[37]** | RMSE | 6.85 | 4.9 | 5.47 | 10.01 | 5.97 | 4.08 | 5.97 |
|  | Score | 75.91 | 29.46 | 35.71 | 90.26 | 54.99 | 31.64 | 54.99 |
| **DGP[38]** | RMSE | 7.03 | 6.17 | 7.26 | 9.86 | 8.00 | 6.29 | 14.35 |
|  | Score | 14212.9 | 5577.33 | 23029.42 | 36310.33 | 15499.20 | 12746.41 | 43083.47 |
| **DAT[39]** | RMSE | 4.57 | 5.35 | 6.73 | 9.25 | 5.18 | 4.08 | 6.71 |
|  | Score | 74 | 91 | 145 | 209 | 99 | 69 | 143 |
| **CaSTAN[40]** | RMSE | - | 4.23 | - | - | - | - | - |
|  | Score | - | 79 | - | - | - | - | - |
| **DiffRUL[41]** | RMSE | 4.86 | 5.81 | 7.95 | 8.48 | - | - | - |
|  | Score | 44.73 | 29.53 | 69.11 | 77.22 | - | - | - |
| **GAF-CNN-Transformer[42]** | RMSE | 5.62 | 5.16 | - | - | - | - | - |
|  | Score | 45 | 39 | - | - | - | - | - |
| **BGT[43]** | RMSE | **4.28** | 6.34 | 5.81 | **7.74** | 6.21 | 4.44 | **5.07** |
|  | Score | 131.69 | 134.99 | 260.01 | 306.26 | 192.52 | 133.58 | 154.61 |
| **OCMM[44]** | RMSE | 7.66 | - | - | 10.03 | - | 8.10 | - |
| | Score | 1.04 | - | - | 1.49 | - | 0.78 | - |
| **DART[39]** | RMSE | 4.57 | **4.16** | **5.18** | 8.15 | **4.62** | **4.08** | 5.94 |
| | Score | 74 | 62 | 93 | 199 | 80 | 69 | 113 |
| **CIDPF** | RMSE | **3.17** | **3.44** | **4.27** | **6.90** | **4.41** | **3.97** | 5.91 |
| | Score | 6654.56 | 3746.80 | 15741.55 | 8778.57 | 2329.85 | 8498.46 | 15005.29 |

## 5. Conclusion

This study proposed the **CIDPF**, a novel approach that bridges the gap between data quality and model efficiency for industrial RUL prediction. By combining causal inference, probabilistic screening, and parameter-efficient transfer learning with LLMs, CIDPF addresses key challenges in high-dimensional and noisy sensor data, achieving SOTA performance on the N-CMAPSS dataset.

CIDPF pioneers the use of causal inference to identify sensor signals with robust and meaningful relationships to RUL, ensuring the retention of only causally relevant data. To further refine the data, a probabilistic GMM-based screening isolates residual noise and anomalies, improving the statistical quality of the training dataset. These steps enable the efficient fine-tuning of pre-trained LLMs, leveraging their powerful sequence modeling capabilities while reducing computational costs and training time. The framework demonstrated robust adaptability to multi-condition and dynamically changing data distributions, making it particularly suitable for real-world industrial applications where data variability and sparsity are challenging.

The results highlight CIDPF's ability to significantly enhance model generalization and accuracy while offering a scalable and resource-efficient solution for industrial prognostics. Beyond achieving superior performances compared to existing methods, CIDPF also reduces reliance on large-scale, high-quality target domain data, enabling practical deployment in cost-sensitive and resource-constrained environments.

Future work will focus on extending CIDPF to cross-domain applications, improving computational efficiency for real-time deployment, and incorporating uncertainty quantification to further enhance prediction reliability. By integrating causal reasoning, probabilistic modeling, and advanced sequence learning, CIDPF sets a foundation for future research at the intersection of predictive maintenance and industrial AI.

## Reference

[1] S. Xiang, Y. Qin, J. Luo, H. Pu, Spatiotemporally multidifferential processing deep neural network and its application to equipment remaining useful life prediction, IEEE Transactions on Industrial Informatics, 18 (2021) 7230-7239.
[2] L. Ren, T. Wang, Z. Jia, F. Li, H. Han, A lightweight and adaptive knowledge distillation framework for remaining useful life prediction, IEEE Transactions on Industrial Informatics, 19 (2022) 9060-9070.
[3] H. Li, Z. Zhang, T. Li, X. Si, A review on physics-informed data-driven remaining useful life prediction: Challenges and opportunities, Mechanical Systems and Signal Processing, 209 (2024) 111120.
[4] L. Zhuang, A. Xu, Y. Wang, Y. Tang, Remaining useful life prediction for two-phase degradation model based on reparameterized inverse Gaussian process, European Journal of Operational Research, 319 (2024) 877-890.
[5] Y. Li, T. Han, T. Xia, Z. Chen, E. Pan, A CM&CP framework with a GIACC method and an ensemble model for remaining useful life prediction, Computers in Industry, 144 (2023) 103794.
[6] A. Vaswani, N. Shazeer, N. Parmar, J. Uszkoreit, L. Jones, A.N. Gomez, Ł. Kaiser, I. Polosukhin, Attention is all you need, Advances in neural information processing systems, 30 (2017).
[7] S. Xiang, P. Li, Y. Huang, J. Luo, Y. Qin, Single gated RNN with differential weighted information storage mechanism and its application to machine RUL prediction, Reliability Engineering & System Safety, 242 (2024) 109741.
[8] Q. Zhang, Q. Liu, Q. Ye, An attention-based temporal convolutional network method for predicting remaining useful life of aero-engine, Engineering Applications of Artificial Intelligence, 127 (2024) 107241.

[9] Z. Xu, V. Selvaraj, S. Min, State identification of a 5-axis ultra-precision CNC machine tool using energy consumption data assisted by multi-output densely connected 1D-CNN model, Journal of Intelligent Manufacturing, 35 (2024) 147-160.
[10] X. Fu, C. Zhang, X. Zhang, H. Sun, A novel GAN architecture reconstructed using BI-LSTM and style transfer for PV temporal dynamics simulation, IEEE Transactions on Sustainable Energy, (2024).
[11] L. Wang, Y. Chen, X. Zhao, J. Xiang, Predictive maintenance scheduling for aircraft engines based on remaining useful life prediction, IEEE Internet of Things Journal, (2024).
[12] G.D. Karatzinis, Y.S. Boutalis, S. Van Vaerenbergh, Aircraft engine remaining useful life prediction: A comparison study of Kernel Adaptive Filtering architectures, Mechanical Systems and Signal Processing, 218 (2024) 111551.
[13] Z. Chen, J. Chan, Large language model in creative work: The role of collaboration modality and user expertise, Management Science, 70 (2024) 9101-9117.
[14] Z. Pang, Y. Luan, J. Chen, T. Li, ParInfoGPT: An LLM-based two-stage framework for reliability assessment of rotating machine under partial information, Reliability Engineering & System Safety, 250 (2024) 110312.
[15] J. Pan, B. Sun, Z. Wu, Z. Yi, Q. Feng, Y. Ren, Z. Wang, Probabilistic remaining useful life prediction without lifetime labels: A Bayesian deep learning and stochastic process fusion method, Reliability Engineering & System Safety, 250 (2024) 110313.
[16] G. Li, L. Yang, C.-G. Lee, X. Wang, M. Rong, A Bayesian deep learning RUL framework integrating epistemic and aleatoric uncertainties, IEEE Transactions on Industrial Electronics, 68 (2020) 8829-8841.
[17] J. Chen, R. Huang, Z. Chen, W. Mao, W. Li, Transfer learning algorithms for bearing remaining useful life prediction: A comprehensive review from an industrial application perspective, Mechanical Systems and Signal Processing, 193 (2023) 110239.
[18] C. Liu, Y. Chen, X. Xu, W. Che, Domain generalization-based damage detection of composite structures powered by structural digital twin, Composites Science and Technology, 258 (2024) 110908.
[19] C. Liu, Y. Chen, X. Xu, Fatigue life prognosis of composite structures using a transferable deep reinforcement learning-based approach, Composite Structures, 353 (2025) 118727.
[20] H. Li, P. Cao, X. Wang, Y. Li, B. Yi, M. Huang, Pre-training enhanced unsupervised contrastive domain adaptation for industrial equipment remaining useful life prediction, Advanced Engineering Informatics, 60 (2024) 102517.
[21] L. Ren, H. Wang, T. Mo, L.T. Yang, A lightweight group transformer-based time series reduction network for edge intelligence and its application in industrial RUL prediction, IEEE Transactions on Neural Networks and Learning Systems, (2024).
[22] N. Houlsby, A. Giurgiu, S. Jastrzebski, B. Morrone, Q. De Laroussilhe, A. Gesmundo, M. Attariyan, S. Gelly, Parameter-efficient transfer learning for NLP, International conference on machine learning, PMLR, 2019, pp. 2790-2799.
[23] Y. Chen, X. Xu, C. Liu, Few-shot meta transfer learning-based damage detection of composite structures, Smart Materials and Structures, 33 (2024) 025027.
[24] C. Liu, Y. Chen, X. Xu, Structural digital Twin for damage detection of CFRP composites using meta transfer Learning-based approach, Expert Systems with Applications, 261 (2025) 125458.
[25] H. Song, M. Kim, D. Park, Y. Shin, J.-G. Lee, Learning from noisy labels with deep neural networks: A survey, IEEE transactions on neural networks and learning systems, 34 (2022) 8135-8153.
[26] C.M. Gilligan-Lee, C. Hart, J. Richens, S. Johri, Leveraging directed causal discovery to detect latent common causes in cause-effect pairs, IEEE Transactions on Neural Networks and Learning Systems, 35 (2022) 4938-4947.
[27] W. Shao, Z. Ge, Z. Song, Semisupervised Bayesian Gaussian mixture models for non-Gaussian soft sensor, IEEE Transactions on Cybernetics, 51 (2019) 3455-3468.
[28] M. Arias Chao, C. Kulkarni, K. Goebel, O. Fink, Aircraft engine run-to-failure dataset under real flight conditions for prognostics and diagnostics, Data, 6 (2021) 5.
[29] J. Runge, P. Nowack, M. Kretschmer, S. Flaxman, D. Sejdinovic, Detecting and quantifying causal associations in large nonlinear time series datasets, Science advances, 5 (2019) eaau4996.
[30] J. Wu, X.-Y. Chen, H. Zhang, L.-D. Xiong, H. Lei, S.-H. Deng, Hyperparameter optimization for machine learning models based on Bayesian optimization, Journal of Electronic Science and Technology, 17 (2019) 26-40.
[31] A. Waswani, N. Shazeer, N. Parmar, J. Uszkoreit, L. Jones, A. Gomez, L. Kaiser, I. Polosukhin, Attention is all you need, NIPS, 2017.
[32] A. Radford, J. Wu, R. Child, D. Luan, D. Amodei, I. Sutskever, Language models are unsupervised multitask learners, OpenAI blog, 1 (2019) 9.
[33] A. Saxena, K. Goebel, D. Simon, N. Eklund, Damage propagation modeling for aircraft engine run-to-failure simulation, 2008 international conference on prognostics and health management, IEEE, 2008, pp. 1-9.
[34] Y. Chen, C. Liu, Remaining Useful Life Prediction: A Study on Multidimensional Industrial Signal Processing and Efficient Transfer Learning Based on Large Language Models, arXiv preprint arXiv:2410.03134, (2024).
[35] E. Frantar, D. Alistarh, Optimal brain compression: A framework for accurate post-training quantization and pruning, Advances in Neural Information Processing Systems, 35 (2022) 4475-4488.
[36] Y. Zhang, H. Wang, Diverse embedding expansion network and low-light cross-modality benchmark for visible-infrared person re-identification, Proceedings of the IEEE/CVF conference on computer vision and pattern recognition, 2023, pp. 2153-2162.
[37] P. Wang, S. Niu, H. Cui, W. Zhang, GPT-based equipment remaining useful life prediction, Proceedings of the ACM Turing Award Celebration Conference-China 2024, 2024, pp. 159-164.
[38] J. Zeng, Z. Liang, A deep Gaussian process approach for predictive maintenance, IEEE Transactions on Reliability, 72 (2022) 916-933.
[39] X. Li, J. Li, L. Zuo, L. Zhu, H.T. Shen, Domain adaptive remaining useful life prediction with transformer, IEEE Transactions on Instrumentation and Measurement, 71 (2022) 1-13.

[40] S. Zheng, J. Liu, Y. Chen, Y. Fan, D. Xu, Causal graph-based spatial–temporal attention network for RUL prediction of complex systems, Computers & Industrial Engineering, (2025) 110892.
[41] W. Wang, H. Song, S. Si, W. Lu, Z. Cai, Data augmentation based on diffusion probabilistic model for remaining useful life estimation of aero-engines, Reliability Engineering & System Safety, 252 (2024) 110394.
[42] X. Yang, X. Gao, H. Zheng, M. Yang, Y. Liu, A hybrid prognosis method based on health indicator and wiener process: The case of multi-sensor monitored aero-engine, Engineering Applications of Artificial Intelligence, 144 (2025) 110099.
[43] F. Xiang, Y. Zhang, S. Zhang, Z. Wang, L. Qiu, J.-H. Choi, Bayesian gated-transformer model for risk-aware prediction of aero-engine remaining useful life, Expert Systems with Applications, 238 (2024) 121859.
[44] J. Wen, J. Ren, Z. Zhao, Z. Zhai, X. Chen, Residual-based adversarial feature decoupling for remaining useful life prediction of aero-engines under variable operating conditions, Expert Systems with Applications, (2024) 124538.